\documentclass{article}
\usepackage{amsmath}
\usepackage{amsfonts}
\usepackage{amssymb}
\usepackage{graphicx}

\input{tcilatex}
\begin{document}

\title{Entropic Dynamics, Time and Quantum Theory\thanks{%
Published in J. Phys. A: Math. Theor. \textbf{44} (2011) 225303.} }
\author{Ariel Caticha \\
%EndAName
{\small Department of Physics, University at Albany-SUNY, }\\
{\small Albany, NY 12222, USA.}}
\date{}
\maketitle

\begin{abstract}
Quantum mechanics is derived as an application of the method of maximum
entropy. No appeal is made to any underlying classical action principle
whether deterministic or stochastic. Instead, the basic assumption is that
in addition to the particles of interest $x$ there exist extra variables $y$
whose entropy $S(x)$ depends on $x$. The Schr\"{o}dinger equation follows
from their coupled dynamics: the entropy $S(x)$ drives the dynamics of the
particles $x$ while they in their turn determine the evolution of $S(x)$. In
this \textquotedblleft entropic dynamics\textquotedblright\ time is
introduced as a device to keep track of change. A welcome feature of such an
entropic time is that it naturally incorporates an arrow of time. Both the
magnitude and the phase of the wave function are given statistical
interpretations: the magnitude gives the distribution of $x$ in agreement
with the usual Born rule and the phase carries information about the entropy 
$S(x)$ of the extra variables. Extending the model to include external
electromagnetic fields yields further insight into the nature of the quantum
phase.
\end{abstract}

\section{Introduction}

The discoveries of black hole entropy \cite{Bekenstein 73} and
thermodynamics \cite{Bardeen 73}, and its connection to quantum theory \cite%
{Hawking 76} suggest a deep connection between the fundamental laws of
physics and information. The details of the connection are not yet known but
one possibility is worth exploring: perhaps what we call physics is the very
useful framework that has been gradually developed to process information
and make inferences about nature. From this perspective one might expect
that the actual rules for processing information---probability theory and
entropic methods---should play central roles in the laws of physics. There
is at least one example where we know this is true: statistical mechanics
and thermodynamics have been derived as an application of the method of
maximum entropy \cite{Jaynes 57}.

Our goal is to derive quantum theory as an application of entropic inference.

One important difference with other approaches that also emphasize notions
of information (see \emph{e.g.}, \cite{Wootters 81}-\cite{Goyal 08}) is the
privileged role we assign to position over and above all other observables.
Our emphasis on position leads to formal similarities with Nelson's
stochastic mechanics \cite{Nelson 66}-\cite{Smolin 06}.

Both the entropic dynamics developed here and Nelson's stochastic mechanics
derive quantum theory as a kind of non-dissipative Brownian motion, but
there are important conceptual differences. First, stochastic mechanics
operates at the ontological level; its goal is to attain a realistic
interpretation of quantum theory as arising from a deeper, possibly
non-local, but essentially classical reality. Entropic dynamics, however,
operates almost completely at the epistemological level; the emphasis is on
making predictions on the basis of limited information. A second difference
is that stochastic mechanics requires two assumptions---the existence of a
universal Brownian motion and that the current velocity is the gradient of
some scalar function---which in entropic dynamics are derived and not merely
postulated. Yet a third difference is that stochastic mechanics is somewhat
closer in spirit to Smoluchowski's approach to the theory of Brownian motion
which involves keeping track of the microscopic details of molecular
collisions through a stochastic Langevin equation and then taking suitable
averages. Entropic dynamics, on the other hand, is closer to the Einstein
approach and focuses on those pieces of information that turn out to be
directly relevant for the prediction of macroscopic effects. The advantage
of the latter approach is the simplicity that arises from not having to keep
track of irrelevant details that are eventually washed out in the averages.

There exist other derivations of the Schr\"{o}dinger that, like the theory
proposed here, do not appeal to the notion of a \textquotedblleft
quantum\textquotedblright\ probability. Some share with Nelson's approach
the foundation of an underlying classical dynamics with an additional
stochastic element. The sub-quantum dynamics is variously described by a
classical action principle, or a Liouville equation, or Newton's law. The
additional stochastic element has been introduced in a variety of ways:
through an extra momentum fluctuation \cite{Hall Reginatto 02}\cite{Olavo 99}%
; a hidden non-equilibrium thermodynamics \cite{Grossing 08}; Brownian
fluctuations caused by energy exchanges with the surrounding vacuum \cite%
{Fritsche 03}; coarse graining an underlying dynamics that is
reparametrization-invariant and ergodic \cite{Elze 02}; tracing out certain
inaccessible degrees of freedom \cite{Wetterich 10}; through explicit
dissipation \cite{tHooft 88}; and also as the statistical mechanics of a
particular class of matrix models \cite{Adler 02}. In contrast, the entropic
dynamics proposed here does not assume any underlying dynamics whether
classical, deterministic, or stochastic. Both quantum theory and its
classical limit are derived as examples of entropic inference.

The statistical model is described in section 2. The basic assumption is
that in addition to the particles of interest there exist some extra
variables. In section 3 we address the basic dynamical question: given that
particles move from an initial point in configuration space to some other
point in its vicinity, where will we expect to find them? When approached as
an example of inference the problem is to select a probability distribution
from within a family specified through appropriate constraints. The answer
is given by the method of maximum entropy (ME) and the most probable new
distribution is that which maximizes an entropy (see \emph{e.g. }\cite%
{Caticha 08b}). The resulting entropic\ dynamics is very simple: except for
fluctuations leading to diffusion the particles tend to drift along the
gradient of the entropy of the extra variables. (Related work on entropic
dynamics appeared in \cite{Caticha 00a}-\cite{Caticha 09}.)

The problem of keeping track of how a succession of small changes builds up
into a large change requires the introduction of a suitable notion of time.
The notions of instant, duration, and the directionality of \emph{entropic }%
time are the subjects of sections 4. Later, in section 8, we discuss the
relation between entropic time and the presumably more fundamental notion of
\textquotedblleft physical\textquotedblright\ time and argue that the latter
is not needed. Our approach is compatible with the theory of \emph{dynamical 
}time advocated by J. Barbour in the context of classical physics (see \emph{%
e.g.} \cite{Barbour 94}).

The dynamics obtained in this way is standard diffusion and is described by
a Fokker-Planck equation (section 5) but quantum mechanics is not just
diffusion. In addition to a probability distribution $\rho (x,t)$ from which
we can generate the magnitude of a wave function we need a second degree of
freedom that would be associated with its phase. The missing ingredient is
supplied in section 6.

Entropic dynamics resembles general relativity in one important respect.
According to general relativity the geometry of space-time dictates how
matter must move, and matter reacts back and dictates how the geometry must
change in response. The lesson of general relativity, if there is one at
all, is that there is no fixed background: space is a dynamical entity. The
situation in entropic dynamics is somewhat analogous: The probability
distributions for the extra variables constitute a statistical manifold, and
it is this space that dictates how the distribution $\rho (x,t)$ diffuses.
To the extent that the statistical manifold is kept frozen one obtains a
fairly standard diffusion. Quantum dynamics arises when we allow the
distribution $\rho (x,t)$ to react back on the statistical manifold. Once
the manifold is promoted to a dynamical entity we have a second degree of
freedom---the entropy of the extra variables---which can be codified into
the phase of a wave function. The dynamics of the manifold is specified,
following Nelson \cite{Nelson 79}, by requiring that a suitable quantity,
which will be called \textquotedblleft energy\textquotedblright , be
conserved and time-reversal invariant (see also \cite{Smolin 06}). This step
completes the derivation of the Schr\"{o}dinger equation.

The analogy to general relativity is further expanded in section 7 with the
introduction of a quantum analogue of the gravitational equivalence
principle. The entropic approach throws new light on old issues such as the
linearity and unitarity of the Schr\"{o}dinger equation, the central role
played by complex numbers, and provides a statistical interpretation not
just of the magnitude of the wave function but also of its phase. In section
9 the model is extended to account for external electromagnetic fields and
the corresponding gauge/phase transformations which provide further insight
into the nature of the quantum phase. Final conclusions are collected in
section 10.

\section{The statistical model}

Consider particles living in flat three-dimensional space. For a single
particle the configuration space $\mathcal{X}$ is Euclidean with metric 
\begin{equation}
\gamma _{ab}=\frac{\delta _{ab}}{\sigma ^{2}}~.  \label{gamma metric}
\end{equation}%
(The reason for the scale factor $\sigma ^{2}$ will become clear once we
generalize to $N$ particles.) Our main assumption is that in addition to the
particles there exist some extra variables that live in a space $\mathcal{Y}$
and are subject to uncertainty. The number and nature of the extra variables 
$y\in \mathcal{Y}$ and the origin of their uncertainty need not be
specified---it is a strength of this formulation that our conclusions hold
irrespective of any assumptions about the $y$ variables. We only need to
assume that their uncertainty depends on the location $x$ of the particles
and this is described by some probability distribution $p(y|x)$. As we shall
see it is the entropy of the distributions $p(y|x)$ that plays a significant
role in defining the dynamics of $x$; the finer details of $p(y|x)$ turn out
to be irrelevant.

For a single particle the \emph{statistical manifold} $\mathcal{M}$ of
distributions $p(y|x)$ is three-dimensional: for each $x$ there is a
corresponding $p(y|x)$. Each distribution $p(y|x)\in \mathcal{M}$ can be
conveniently labeled by its corresponding $x$ so that the label $x$ denotes
both a regular point in the configuration space $\mathcal{X}$ and also its
corresponding \textquotedblleft point\textquotedblright\ in the statistical
manifold $\mathcal{M}$. For later reference, the entropy $S(x)$ of $p(y|x)$
relative to an underlying measure $q(y)$ of the extra-variable space $%
\mathcal{Y}$ is\footnote{%
This is a multidimensional integral over all $y$ variables; for simplicity
we write $dy$ instead of $d^{n}y$.} 
\begin{equation}
S(x)=-\int dy\,p(y|x)\log \frac{p(y|x)}{q(y)}~.  \label{entropy a}
\end{equation}%
This entropy $S(x)$ is a natural scalar field on both the configuration
space $\mathcal{X}$ and the statistical manifold $\mathcal{M}$.

The peculiar features of quantum mechanics such as non-local correlations
and entanglement will arise naturally provided the theory for $N$ particles
is formulated on the $3N$-dimensional configuration space $\mathcal{X}_{N}$.
Accordingly, to complete the specification of the model we need to describe $%
\mathcal{X}_{N}$ and its corresponding statistical manifold $\mathcal{M}_{N}$%
. The generalization is straightforward. For $N$ particles the
extra-variable distributions are $p(y|x)$ where now the position $x\in 
\mathcal{X}_{N}$ is given by $x^{A}$ and the index $A$ now takes $3N$
values. More explicitly $x=(x^{a_{1}},x^{a_{2}}\ldots )$ where $a_{1}=1,2,3$
denotes the first particle, $a_{2}=4,5,6$ denotes the second particle, and
so on. The $3N$-dimensional configuration space $\mathcal{X}_{N}$ remains
flat but it is not, in general, isotropic. For example, for $N=2$ particles
the metric, written in block matrix form, is 
\begin{equation}
\gamma _{AB}=%
\begin{bmatrix}
\delta _{a_{1}b_{1}}/\sigma _{1}^{2} & 0 \\ 
0 & \delta _{a_{2}b_{2}}/\sigma _{2}^{2}%
\end{bmatrix}%
~.~  \label{gamma AB}
\end{equation}%
We shall later see that this choice of an anisotropic configuration space
leads to a theory of particles with different masses. For particles that are
identical the appropriate configuration space is isotropic with $\sigma
_{1}=\sigma _{2}=\ldots =\sigma $.

To summarize, the first basic assumption is the existence of some extra
variables $y$ subject to an $x$-dependent uncertainty described by some
unspecified distributions $p(y|x)$. The statistical manifold $\mathcal{M}%
_{N} $ and the entropy field $S(x)$ are convenient inference tools
introduced to explore the implications of this assumption.

\section{Entropic dynamics}

The second basic assumption is that small changes from one state to another
are possible and do, in fact, happen. We do not explain why they happen but,
given the information that changes occur, our problem is to venture a guess
about what changes to expect. Large changes are assumed to result from the
accumulation of many small changes.

Consider a single particle (the generalization to several particles is
immediate) that moves away from an initial position $x$ to an unknown final
position $x^{\prime }$. All we know about $x^{\prime }$ is that it is near $%
x $. What can we say about $x^{\prime }$? Since $x$ and $x^{\prime }$
represent probability distributions we see that this is precisely the kind
of problem the method of maximum entropy (ME) has been designed to solve,
namely, to update from a prior distribution to a posterior distribution
selected from within a specified set.\footnote{%
The notion of updating from a prior to a posterior distribution is standard
in Bayesian inference; it is also appropriate in the context of entropic
inference. For a detailed account of the use of relative entropy as a tool
for updating which includes Bayesian updating as a special case see \cite%
{Caticha 08b}.} As in all ME problems success hinges on appropriate choices
of the entropy, prior distribution, and constraints.

Since neither the new $x^{\prime }$ nor the new extra variables $y^{\prime }$
are known what we want is the joint distribution $P(x^{\prime },y^{\prime
}|x)$ and the relevant space is $\mathcal{X\times Y}$. To find it maximize
the appropriate (relative) entropy, 
\begin{equation}
\mathcal{S}[P,Q]=-\int dx^{\prime }dy^{\prime }\,P(x^{\prime },y^{\prime
}|x)\log \frac{P(x^{\prime },y^{\prime }|x)}{Q(x^{\prime },y^{\prime }|x)}~.
\label{Sppi}
\end{equation}%
The relevant information is introduced through the prior $Q(x^{\prime
},y^{\prime }|x)$ and the constraints that specify the family of acceptable
posteriors $P(x^{\prime },y^{\prime }|x)$.

We select a prior that represents a state of extreme ignorance: the relation
between $x^{\prime }$ and $y^{\prime }$ is not known; knowledge of $%
x^{\prime }$ tells us nothing about $y^{\prime }$ and vice versa. Such
ignorance is represented by a product, $Q(x^{\prime },y^{\prime
}|x)=Q(x^{\prime }|x)Q(y^{\prime }|x)$. Furthermore we take the
distributions $Q(y^{\prime }|x)dy^{\prime }$ and $Q(x^{\prime
}|x)d^{3}x^{\prime }$ to be uniform, that is, proportional to the respective
volume elements which are respectively given by $dv_{x}=\gamma ^{1/2}d^{3}x$
[where $\gamma =\det \gamma _{ab}$, see eq.(\ref{gamma metric})] and by $%
dv_{y}=q(y)dy$ where the measure $q(y)$ need not be specified further.
Therefore, since proportionality constants are not essential here, the joint
prior is 
\begin{equation}
Q(x^{\prime },y^{\prime }|x)=\gamma ^{1/2}q(y)~.  \label{prior}
\end{equation}

Next we specify the constraints. Write the posterior as 
\begin{equation}
P(x^{\prime },y^{\prime }|x)=P(x^{\prime }|x)P(y^{\prime }|x^{\prime },x)
\end{equation}
and consider the two factors separately. First we require that $x^{\prime }$
and $y^{\prime }$ be related to each other in a very specific way, namely
that $P(y^{\prime }|x^{\prime },x)=p(y^{\prime }|x^{\prime })\in \mathcal{M}$%
---the uncertainty in $y^{\prime }$ depends only on $x^{\prime }$, and not
on previous positions $x$. Therefore, our first constraint is that the joint
posterior be of the form%
\begin{equation}
P(x^{\prime },y^{\prime }|x)=P(x^{\prime }|x)p(y^{\prime }|x^{\prime })~.
\label{constraint p}
\end{equation}%
The second constraint concerns the factor $P(x^{\prime }|x)$ and represents
the fact that actual physical changes do not happen discontinuously: we
require that $x^{\prime }$ be an infinitesimally short distance away from $x$%
. Let $x^{\prime a}=x^{a}+\Delta x^{a}$. We require that the expectation 
\begin{equation}
\left\langle \Delta \ell ^{2}(x^{\prime },x)\right\rangle =\left\langle
\gamma _{ab}\Delta x^{a}\Delta x^{b}\right\rangle =\Delta \bar{\ell}^{2}~
\label{short step}
\end{equation}%
be some small but for now unspecified numerical value $\Delta \bar{\ell}^{2}$
which could in principle depend on $x$.

Having specified the prior and the constraints the ME method takes over.
Substituting the prior (\ref{prior}) and the constraint (\ref{constraint p})
into the joint entropy (\ref{Sppi}) gives 
\begin{equation}
\mathcal{S}[P,Q]=-\int dx^{\prime }\,P(x^{\prime }|x)\log \frac{P(x^{\prime
}|x)}{\gamma ^{1/2}}+\int dx^{\prime }\,P(x^{\prime }|x)S(x^{\prime })~,
\label{Sppi b}
\end{equation}%
where $S(x)$ is given in eq.(\ref{entropy a}). Next we vary $P(x^{\prime
}|x) $ to maximize $\mathcal{S}[P,\pi ]$ subject to (\ref{short step}) and
normalization. The result is 
\begin{equation}
P(x^{\prime }|x)=\frac{1}{\zeta (x,\alpha )}e^{S(x^{\prime })-\frac{1}{2}%
\alpha (x)\Delta \ell ^{2}(x^{\prime },x)}~,  \label{Prob xp/x}
\end{equation}%
where 
\begin{equation}
\zeta (x,\alpha )=\int dx^{\prime }\,e^{S(x^{\prime })-\frac{1}{2}\alpha
(x)\Delta \ell ^{2}(x^{\prime },x)}~,  \label{zeta}
\end{equation}%
and the Lagrange multiplier $\alpha (x)$ is determined from the constraint
eq.(\ref{short step}), 
\begin{equation}
\frac{\partial }{\partial \alpha }\log \zeta (x,\alpha )=-\frac{1}{2}\Delta 
\bar{\ell}^{2}~.
\end{equation}%
The distribution (\ref{Prob xp/x}) is not merely a local maximum or a
stationary point, it yields the absolute maximum of the relative entropy $%
\mathcal{S}[P,Q]$ subject to the constraints (\ref{constraint p}) and (\ref%
{short step}). The proof (see the appendix) follows the standard argument
originally due to Gibbs \cite{Jaynes 03}.

The probability of a step from $x$ to $x^{\prime }$, eq.(\ref{Prob xp/x}),
represents a compromise between three conflicting tendencies. One, which can
be traced to the uniform prior $Q(x^{\prime }|x)=\gamma ^{1/2}$ and is
represented by the first integral in (\ref{Sppi b}), is to make $P(x^{\prime
}|x)$ spread as uniformly as possible. Another, induced by the second
integral in (\ref{Sppi b}), contributes the entropy term in the exponent of $%
P(x^{\prime }|x)$ and favors a single giant step to the distribution $%
p(y^{\prime }|x^{\prime })$ that maximizes the entropy $S(x^{\prime })$. And
last, the constraint on $\left\langle \Delta \ell ^{2}\right\rangle $ leads
to the $\Delta \ell ^{2}(x^{\prime },x)$ term in the exponent of $%
P(x^{\prime }|x)$ and favors values of $x^{\prime }$ that are close to $x$.
Large $\alpha $ means short steps. The compromise in eq.(\ref{Prob xp/x})
leads to short steps in essentially random directions with a small
anisotropic bias along the entropy gradient.

For large $\alpha $ let $x^{\prime a}=x^{a}+\Delta x^{a}$. Expanding the
exponent in (\ref{Prob xp/x}) about its maximum gives 
\begin{equation}
P(x^{\prime }|x)\approx \frac{1}{Z(x)}\exp \left[ -\frac{\alpha (x)}{2\sigma
^{2}}\delta _{ab}\left( \Delta x^{a}-\Delta \bar{x}^{a}\right) \left( \Delta
x^{b}-\Delta \bar{x}^{b}\right) \right] ,  \label{Prob xp/x b}
\end{equation}%
where factors independent of $x^{\prime }$ have been absorbed into a new
normalization $Z(x)$. The displacement $\Delta x^{a}$ can be expressed as
the expected drift plus a fluctuation,%
\begin{equation}
\Delta x^{a}=\Delta \bar{x}^{a}+\Delta w^{a}~,  \label{Delta x}
\end{equation}%
where%
\begin{equation}
\left\langle \Delta x^{a}\right\rangle =\Delta \bar{x}^{a}=\frac{\sigma ^{2}%
}{\alpha (x)}\delta ^{ab}\partial _{b}S(x)~,  \label{ED drift}
\end{equation}%
\begin{equation}
\left\langle \Delta w^{a}\right\rangle =0\quad \text{and}\quad \left\langle
\Delta w^{a}\Delta w^{b}\right\rangle =\frac{\sigma ^{2}}{\alpha (x)}\delta
^{ab}~.  \label{ED fluctuations}
\end{equation}%
The particle tends to move along the entropy gradient. Note that as $\alpha
\rightarrow \infty $ the steps get correspondingly smaller but the
fluctuations become dominant: the drift is $\Delta \bar{x}\sim O(\alpha
^{-1})$ while the fluctuations are much larger $\Delta w\sim O(\alpha
^{-1/2})$. This implies that as $\alpha \rightarrow \infty $ the trajectory
is continuous but not differentiable---just as in Brownian motion.

We can now return to the unfinished business of choosing $\Delta \bar{\ell}%
^{2}$ in eq.(\ref{short step}) which is equivalent to choosing the
multiplier $\alpha (x)$. We invoke a symmetry argument. We just saw that in
the limit of infinitesimally short steps the relevant dynamics is dominated
by the fluctuations $\Delta w$. In order that the dynamics reflect the
translational symmetry of the configuration space $\mathcal{X}$ we choose $%
\alpha (x)$ so that the fluctuations $\left\langle \Delta w^{a}\Delta
w^{b}\right\rangle $ in eq.(\ref{ED fluctuations}) be independent of $x$.
Therefore $\alpha (x)=\func{constant}$.

\section{Entropic time}

Our goal is to derive laws of physics as an application of inference methods
but the latter make no reference to time so additional assumptions are
needed. \emph{The foundation to any notion of time is dynamics.} We
introduce time as a convenient book-keeping device to keep track of the
accumulation of small changes.

In this section we show how a dynamics driven by entropy naturally leads to
an \textquotedblleft entropic\textquotedblright\ notion of time. Our task
here is to develop a model that includes (a) something one might identify as
an \textquotedblleft instant\textquotedblright , (b) a sense in which these
instants can be \textquotedblleft ordered\textquotedblright , (c) a
convenient concept of \textquotedblleft duration\textquotedblright\
measuring the separation between instants. A welcome bonus is that the model
incorporates an intrinsic directionality---an evolution from past instants
towards future instants. Thus, an arrow of time does not have to be
externally imposed but is generated automatically. This set of concepts
constitutes what we will call \textquotedblleft entropic
time\textquotedblright .

Important questions such as the relation between entropic time, in which
instants are ordered through the sequence of inference steps, and an
externally imposed structure of a presumably \textquotedblleft
physical\textquotedblright\ time will be discussed later (section 8) after
the dynamics has been more fully developed.

\subsection{Time as a sequence of instants}

In entropic dynamics change is given, at least for infinitesimally short
steps, by the transition probability $P(x^{\prime }|x)$ in eq.(\ref{Prob
xp/x b}). For finite steps the relevant piece of information is that large
changes occur \emph{only} as the result of a continuous succession of very
many small changes.

Consider the $n$th step. In general we will be uncertain about both its
initial and the final positions, $x$ and $x^{\prime }$. This means we must
deal with the joint probability $P(x^{\prime },x)$. Using $P(x^{\prime
},x)=P(x^{\prime }|x)P(x)$ and integrating over $x$, we get 
\begin{equation}
P(x^{\prime })=\tint dx\,P(x^{\prime }|x)P(x)~.  \label{CK a}
\end{equation}%
It is important to emphasize that this equation is a direct consequence of
the laws of probability---no assumptions of a physical nature have been
made. However, if $P(x)$ happens to be the probability of different values
of $x$ \emph{at a given instant of entropic time }$t$, then it is tempting
to interpret $P(x^{\prime })$ as the probability of values of $x^{\prime }$
at a \textquotedblleft later\textquotedblright\ instant of entropic time $%
t^{\prime }=t+\Delta t$. Accordingly, we write $P(x)=\rho (x,t)$ and $%
P(x^{\prime })=\rho (x^{\prime },t^{\prime })$ so that 
\begin{equation}
\rho (x^{\prime },t^{\prime })=\tint dx\,P(x^{\prime }|x)\rho (x,t)
\label{CK b}
\end{equation}%
Nothing in the laws of probability that led to eq.(\ref{CK a}) forces this
interpretation on us---this is an independent assumption about what
constitutes time in our model. We use eq.(\ref{CK b}) to define what we mean
by an instant:\emph{\ if the distribution }$\rho (x,t)$\emph{\ refers to an
\textquotedblleft initial\textquotedblright\ instant, then the distribution }%
$\rho (x^{\prime },t^{\prime })$\emph{\ defines what we mean by the
\textquotedblleft next\textquotedblright\ instant. }Thus, eq.(\ref{CK b})
allows entropic time to be constructed, step by step, as a succession of
instants.\emph{\ }

\subsection{Duration: a convenient time scale}

Having introduced the notion of successive instants we now have to specify
the interval $\Delta t$ between them. Successive instants are connected
through the transition probability $P(x^{\prime }|x)$. Specifying the
interval of time $\Delta t$ between successive instants amounts to tuning
the steps or, equivalently, the multiplier $\alpha (x,t)$. To model a time
that, like Newtonian time, flows \textquotedblleft
equably\textquotedblright\ everywhere, that is, at the same rate at all
places and times we define $\Delta t$ as being independent of $x$, and such
that every $\Delta t$ is as long as the previous one. Inspection of the
actual dynamics as given in eq.(\ref{Prob xp/x b}-\ref{ED fluctuations})
shows that this is achieved if we choose $\alpha (x,t)$ so that 
\begin{equation}
\alpha (x,t)=\frac{\tau }{\Delta t}=\func{constant}~,  \label{alpha}
\end{equation}%
where $\tau $ is a constant introduced so that $\Delta t$ has units of time.
As already anticipated in the previous section, it is the translational
symmetry of the configuration space $\mathcal{X}$ expressed as the
\textquotedblleft equable\textquotedblright\ flow of time that leads us to
impose uniformity on the expected step sizes $\Delta \bar{\ell}$ and the
corresponding multipliers $\alpha $. This completes the implementation of
entropic time. In the end, however, the only justification for any
definition of duration is that it simplifies the description of motion, and
indeed, the transition probability in eq.(\ref{Prob xp/x b}) becomes 
\begin{equation}
P(x^{\prime }|x)\approx \frac{1}{Z(x)}\exp \left[ -\frac{\tau }{2\sigma
^{2}\Delta t}\delta _{ab}\left( \Delta x^{a}-\Delta \bar{x}^{a}\right)
\left( \Delta x^{b}-\Delta \bar{x}^{b}\right) \right] ~,
\label{Prob xp/x bb}
\end{equation}%
which we recognize as a standard Wiener process. A displacement $\Delta
x=x^{\prime }-x$ is given by

\begin{equation}
\Delta x^{a}=b^{a}(x)\Delta t+\Delta w^{a}~,  \label{Delta x b}
\end{equation}%
where the drift velocity $b^{a}(x)$ and the fluctuation $\Delta w^{a}$ are 
\begin{equation}
\langle \Delta x^{a}\rangle =b^{a}\Delta t\quad \text{with}\quad b^{a}(x)=%
\frac{\sigma ^{2}}{\tau }\,\delta ^{ab}\partial _{b}S(x)~,
\label{drift future}
\end{equation}%
\begin{equation}
\left\langle \Delta w^{a}\right\rangle =0\quad \text{and}\quad \left\langle
\Delta w^{a}\Delta w^{b}\right\rangle =\frac{\sigma ^{2}}{\tau }\Delta
t\,\delta ^{ab}~.  \label{fluc}
\end{equation}%
The constant $\sigma ^{2}/2\tau $ plays the role of the diffusion constant
in Brownian motion. The formal similarity to Nelson's stochastic mechanics 
\cite{Nelson 66} is evident. An important difference concerns the expression
of the drift velocity as the gradient of a scalar function: unlike
stochastic mechanics, here eq.(\ref{drift future}) has been derived rather
than postulated, and $S(x)$ is not merely an uninterpreted auxiliary scalar
function---it turns out to be the entropy of the $y$ variables.

\subsection{The directionality of entropic time}

Time constructed according to eq.(\ref{CK b}) is remarkable in yet another
respect: the inference implied by $P(x^{\prime }|x)$ in eq.(\ref{Prob xp/x b}%
) incorporates an intrinsic directionality in entropic time: there is an
absolute sense in which $\rho (x,t)$\ is prior and $\rho (x^{\prime
},t^{\prime })$\ is posterior.

Suppose we wanted to find a time-reversed evolution. We would write 
\begin{equation}
\rho (x,t)=\tint dx^{\prime }\,P(x|x^{\prime })\rho (x^{\prime },t^{\prime
})\,.
\end{equation}%
This is perfectly legitimate but in order to be correct $P(x|x^{\prime })$
cannot be obtained from eq.(\ref{Prob xp/x b}) by merely exchanging $x$ and $%
x^{\prime }$. According to the rules of probability theory $P(x|x^{\prime })$
is related to eq.(\ref{Prob xp/x b}) by Bayes' theorem, 
\begin{equation}
P(x|x^{\prime })=\frac{P(x)}{P(x^{\prime })}P(x^{\prime }|x)~.  \label{bt1}
\end{equation}%
In other words, one of the two transition probabilities, either $%
P(x|x^{\prime })$ or $P(x|x^{\prime })$, \emph{but not both}, can be given
by the maximum entropy distribution eq.(\ref{Prob xp/x b}). The other is
related to it by Bayes' theorem. I hesitate to say that this is what breaks
the time-reversal symmetry because the symmetry was never there in the first
place. There is no symmetry between prior and posterior; there is no
symmetry between the inferential past and the inferential future.

An interesting consequence of the time asymmetry is that the mean velocities
towards the future and from the past do not coincide. Let us be more
specific. Equation (\ref{drift future}) gives the \emph{mean velocity to the
future} or \emph{future drift}, 
\begin{eqnarray}
b^{a}(x) &=&\lim_{\Delta t\rightarrow 0^{+}}\frac{\left\langle
x^{a}(t+\Delta t)\right\rangle _{x(t)}-x^{a}(t)}{\Delta t}  \notag \\
&=&\lim_{\Delta t\rightarrow 0^{+}}\frac{1}{\Delta t}\tint dx^{\prime
}\,P(x^{\prime }|x)\Delta x^{a}  \label{drift future b}
\end{eqnarray}%
where $x=x(t)$, $x^{\prime }=x(t+\Delta t)$, and $\Delta x^{a}=x^{\prime
a}-x^{a}$. Note that the expectation in (\ref{drift future b}) is
conditional on the earlier position $x=x(t)$. One can also define a \emph{%
mean velocity from the past} or \emph{past drift}, 
\begin{equation}
b_{\ast }^{a}(x)=\lim_{\Delta t\rightarrow 0^{+}}\frac{x^{a}(t)-\left\langle
x^{a}(t-\Delta t)\right\rangle _{x(t)}}{\Delta t}  \label{drift past}
\end{equation}%
where the expectation is conditional on the later position $x=x(t)$.
Shifting the time by $\Delta t$, $b_{\ast }^{a}$ can be equivalently written
as 
\begin{eqnarray}
b_{\ast }^{a}(x^{\prime }) &=&\lim_{\Delta t\rightarrow 0^{+}}\frac{%
x^{a}(t+\Delta t)-\left\langle x^{a}(t)\right\rangle _{x(t+\Delta t)}}{%
\Delta t}  \notag \\
&=&\lim_{\Delta t\rightarrow 0^{+}}\frac{1}{\Delta t}\tint dx\,P(x|x^{\prime
})\Delta x^{a}~,  \label{drift past b}
\end{eqnarray}%
with the same definition of $\Delta x^{a}$ as in eq.(\ref{drift future b}).

The two mean velocities, to the future $b^{a}$, and from the past $b_{\ast
}^{a}$, do not coincide. The connection between them is well known \cite%
{Nelson 66}\cite{Nelson 85},

\begin{equation}
b_{\ast }^{a}(x,t)=b^{a}(x)-\frac{\sigma ^{2}}{\tau }\partial ^{a}\log \rho
(x,t)~,  \label{drift past c}
\end{equation}%
where\footnote{%
From now on indices are raised and lowered with the Euclidean metric $\delta
_{ab}$.} $\partial ^{a}=\delta ^{ab}\partial _{b}$ and $\rho (x,t)=P(x)$.
What might not be widely appreciated is that eq.(\ref{drift past c}) is a
straightforward consequence of Bayes' theorem, eq. (\ref{bt1}). (For a
related idea see \cite{Jaynes 89}.) To derive eq.(\ref{drift past c}) expand 
$P(x^{\prime })$ about $x$ in (\ref{bt1}) to get 
\begin{equation}
P(x|x^{\prime })=\left[ 1-\frac{\partial \log \rho (x,t)}{\partial x^{b}}%
\,\Delta x^{b}+\ldots \right] P(x^{\prime }|x)\,.  \label{bt2}
\end{equation}%
Multiply $b_{\ast }^{a}(x^{\prime })$ in eq.(\ref{drift past b}) by a smooth
test function $f(x^{\prime })$ and integrate, 
\begin{equation}
\tint dx^{\prime }\,b_{\ast }^{a}(x^{\prime })f(x^{\prime })=\frac{1}{\Delta
t}\tint dx^{\prime }\tint dx\,P(x|x^{\prime })\Delta x^{a}f(x^{\prime })~.
\end{equation}%
(The limit $\Delta t\rightarrow 0^{+}$ is understood.) On the right hand
side expand $f(x^{\prime })$ about $x$ and use (\ref{bt2}), 
\begin{equation}
\frac{1}{\Delta t}\tint dx^{\prime }\tint dx\,P(x^{\prime }|x)\left[ \Delta
x^{a}f(x)-\Delta x^{a}\Delta x^{b}\frac{\partial \log \rho (x,t)}{\partial
x^{b}}f(x)+\Delta x^{a}\Delta x^{c}\frac{\partial f}{\partial x^{c}}+\ldots %
\right] .  \label{bt3}
\end{equation}%
Next interchange the orders of integration and take $\Delta t\rightarrow
0^{+}$ using eq.(\ref{fluc}), 
\begin{equation}
\langle \Delta x^{a}\Delta x^{b}\rangle =\langle \Delta w^{a}\Delta
w^{b}\rangle =\frac{\sigma ^{2}}{\tau }\Delta t\,\delta ^{ab}~.
\label{fluc b}
\end{equation}%
On integration by parts the third term of (\ref{bt3}) vanishes and we get 
\begin{equation}
\tint dx\,b_{\ast }^{a}(x)f(x)=\tint dx\,\left[ b^{a}(x)-\frac{\sigma ^{2}}{%
\tau }\delta ^{ab}\partial _{b}\log \rho (x,t)\right] f(x)\,,
\end{equation}%
Since $f(x)$ is arbitrary we get (\ref{drift past c}).

The puzzle of the arrow of time has a long history (see \emph{e.g.} \cite%
{Price96 Zeh01}). The standard question is how can an arrow of time be
derived from underlying laws of nature that are symmetric? Entropic dynamics
offers a new perspective because it does not assume any underlying laws of
nature -- whether they be symmetric or not -- and its goal is not to explain
the asymmetry between past and future. The asymmetry is the inevitable
consequence of entropic inference. From the point of view of entropic
dynamics the challenge does not consist in explaining the arrow of time, but
rather in explaining how it comes about that despite the arrow of time some
laws of physics turn out to be reversible. Indeed, even when the derived
laws of physics---in our case, the Schr\"{o}dinger equation---turns out to
be fully time-reversible, \emph{entropic time itself only f{}lows forward}.

\section{Accumulating changes: the Fokker-Planck equation}

Time has been introduced as a useful device to keep track of the
accumulation of small changes. The technique to do this is well known from
diffusion theory \cite{Chandrasekhar 43}. The equation of evolution for the
distribution $\rho (x,t)$, derived from eq.(\ref{CK b}) together with (\ref%
{Delta x b})-(\ref{fluc}), is a Fokker-Planck equation (FP), 
\begin{equation}
\partial _{t}\rho =-\partial _{a}\left( b^{a}\rho \right) +\frac{\sigma ^{2}%
}{2\tau }\nabla ^{2}\rho ~,  \label{FP a}
\end{equation}%
where $\partial _{a}=\partial /\partial x^{a}$ and $\nabla ^{2}=\delta
^{ab}\partial ^{2}/\partial x^{a}\partial x^{b}$. The FP equation can be
rewritten as a continuity equation, 
\begin{equation}
\partial _{t}\rho =-\partial _{a}\left( v^{a}\rho \right) ~,  \label{FP b}
\end{equation}%
where the velocity of the probability flow or \emph{current velocity} is%
\begin{equation}
v^{a}=b^{a}-\frac{\sigma ^{2}}{2\tau }\delta ^{ab}\frac{\partial _{a}\rho }{%
\rho }~.
\end{equation}%
It is convenient to introduce the \emph{osmotic velocity} 
\begin{equation}
u^{a}\overset{\text{def}}{=}-\frac{\sigma ^{2}}{2\tau }\partial ^{a}\log
\rho ~.  \label{osmo}
\end{equation}%
\emph{\ }Its interpretation follows from $v^{a}=b^{a}+u^{a}$. The drift $%
b^{a}$, eq.(\ref{drift future}), represents the tendency of the probability $%
\rho $ to flow up the entropy gradient while $u^{a}$ represents the tendency
to flow down the density gradient. The situation is analogous to Brownian
motion where the drift velocity is the response to the gradient of an
external potential, while $u^{a}$ is a response to the gradient of
concentration or chemical potential---the so-called\emph{\ osmotic force}.%
\footnote{%
The definition of osmotic velocity adopted by Nelson \cite{Nelson 66} and
other authors differs from ours by a sign. Nelson takes the osmotic velocity
to be the velocity imparted by a force applied externally in order to
balance the osmotic force (due to concentration gradients) and attain
equilibrium. For us the osmotic velocity is the velocity imparted by the
osmotic force itself.} The osmotic contribution to the probability flow is
the actual diffusion current, 
\begin{equation}
\rho u^{a}=-\frac{\sigma ^{2}}{2\tau }\partial ^{a}\rho ~,
\end{equation}%
which can be recognized as Fick's law, with a diffusion coefficient given by 
$\sigma ^{2}/2\tau $.

Since both the future drift $b^{a}$ and the osmotic velocity $u^{a}$ are
gradients, it follows that the current velocity is a gradient too. For later
reference, from (\ref{drift future}) and (\ref{osmo}), 
\begin{equation}
v^{a}=\frac{\sigma ^{2}}{\tau }\,\partial ^{a}\phi ~,  \label{curr}
\end{equation}%
where $\phi (x,t)=S(x)-\log \rho ^{1/2}(x,t)$.

With these results entropic dynamics reaches a certain level of completion:
We figured out what small changes to expect---they are given by $P(x^{\prime
}|x)$---and time was introduced to keep track of how these small changes
accumulate; the net result is diffusion according to the FP equation (\ref%
{FP a}).

\section{Manifold dynamics}

But quantum mechanics is not \emph{just} diffusion. The description so far
has led us to the density $\rho (x,t)$ as the important dynamical object but
to construct a wave function, $\Psi =\rho ^{1/2}e^{i\phi }$, we need a
second degree of freedom, the phase $\phi $. The problem is that as long as
the geometry of the statistical manifold $\mathcal{M}$ is rigidly fixed
there is no logical room for additional degrees of freedom. The natural
solution is to remove this constraint. We allow the manifold $\mathcal{M}$
to participate in the dynamics and the entropy of the $y$ variables becomes
a time-dependent field, $S(x,t)$. We can take $S(x,t)$ to be the new
independent degree of freedom but eq.(\ref{curr}) suggests that a more
convenient and yet equivalent choice is%
\begin{equation}
\phi (x,t)=S(x,t)-\log \rho ^{1/2}(x,t)~.  \label{phase}
\end{equation}%
Thus the dynamics will consist of the coupled evolution of $\rho (x,t)$ and $%
\phi (x,t)$.

\subsection{Conservative diffusion}

To specify the dynamics of the manifold $\mathcal{M}$ we follow Nelson \cite%
{Nelson 79} and impose that the dynamics be \textquotedblleft
conservative,\textquotedblright\ that is, one requires the conservation of a
certain functional $E[\rho ,S]$ of $\rho (x,t)$ and $S(x,t)$.

Requiring that some \textquotedblleft energy\textquotedblright\ $E[\rho ,S]$%
\ be conserved may seem natural because it clearly represents relevant
information but it is an assumption that cries out for a deeper
justification. Normally energy is whatever happens to be conserved as a
result of invariance under translations in time. But our dynamics has not
been defined yet; what, then, is \textquotedblleft energy\textquotedblright\
and why should it be conserved in the first place? This is a question best
left for the future. At this early stage, for the purpose of deriving a
non-relativistic model, we just propose an intuitively reasonable conserved
energy and proceed.

The particular form of $E[\rho ,S]$ is chosen to be a local functional that
is invariant under time reversal and we require that the velocities enter in
rotationally invariant terms \cite{Smolin 06}. Under time reversal $%
t\rightarrow -t$ we have 
\begin{equation}
b^{a}\rightarrow -b_{\ast }^{a}\,,\quad v^{a}\rightarrow -v^{a}\,,\quad
u^{a}\rightarrow u^{a}\,,  \label{t reversal}
\end{equation}%
In the low velocity limit this means we need only include velocity terms in $%
v^{2}$ and $u^{2}$. The proposed energy functional is 
\begin{equation}
E[\rho ,S]=\int d^{3}x\,\rho (x,t)\left( A\gamma _{ab}v^{a}v^{b}+B\gamma
_{ab}u^{a}u^{b}+V(x)\right) ~,  \label{energy a}
\end{equation}%
where $A$ and $B$ are constants, $\gamma _{ab}$ is given by (\ref{gamma
metric}), and $V(x)$ represents an external potential. If $E$ has units of
energy then $A/\sigma ^{2}$ and $B/\sigma ^{2}$ have units of mass. Let us
define new constants 
\begin{equation}
m=\frac{2A}{\sigma ^{2}}\quad \text{and\quad }\mu =\frac{2B}{\sigma ^{2}}~,
\end{equation}%
which we will call the \textquotedblleft current mass\textquotedblright\ and
the \textquotedblleft osmotic mass\textquotedblright . Then the energy
functional is 
\begin{equation}
E[\rho ,S]=\int d^{3}x\,\rho (x,t)\left( \frac{1}{2}mv^{2}+\frac{1}{2}\mu
u^{2}+V(x)\right) ~.  \label{energy b}
\end{equation}%
It is further convenient to combine the constant $\tau $, which sets the
units of time, with $A$ into yet a new constant $\eta $, 
\begin{equation}
\eta =\frac{2A}{\tau }\quad \text{so that}\quad \frac{\sigma ^{2}}{\tau }=%
\frac{\eta }{m}~.  \label{eta}
\end{equation}%
$\eta $ relates the units of mass or energy with those of time. Then the
current and osmotic velocities, eqs.(\ref{curr}) and (\ref{osmo}) are 
\begin{equation}
mv_{a}=\eta \,\partial _{a}\phi \quad \text{and\quad }mu_{a}=-\eta \partial
_{a}\log \rho ^{1/2}~,  \label{curr osmo}
\end{equation}%
and the energy (\ref{energy b}) becomes 
\begin{equation}
E[\rho ,S]=\int dx\,\rho \left( \frac{\eta ^{2}}{2m}(\partial _{a}\phi )^{2}+%
\frac{\mu \eta ^{2}}{8m^{2}}(\partial _{a}\log \rho )^{2}+V\right) ~.
\label{energy c}
\end{equation}

When the potential is static, $\dot{V}=0$, energy is conserved, $\dot{E}=0$.
Otherwise we impose that energy increase at the rate 
\begin{equation}
\dot{E}=\int dx\,\rho \dot{V}~.
\end{equation}%
Next take the time derivative of (\ref{energy c}). After integrating by
parts, and using eqs.(\ref{FP b}) and (\ref{curr}), 
\begin{equation}
\dot{\rho}=-\partial _{a}\left( \rho v^{a}\right) =-\frac{\eta }{m}\partial
^{a}\left( \rho \partial _{a}\phi \right) =-\frac{\eta }{m}\left( \partial
^{a}\rho \partial _{a}\phi +\rho \nabla ^{2}\phi \right) ~,  \label{SEa}
\end{equation}%
we get 
\begin{equation}
\dot{E}-\int dx\,\rho \dot{V}=\int dx\,\dot{\rho}\left[ \eta \dot{\phi}+%
\frac{\eta ^{2}}{2m}(\partial _{a}\phi )^{2}+V-\frac{\mu \eta ^{2}}{2m^{2}}%
\frac{\nabla ^{2}\rho ^{1/2}}{\rho ^{1/2}}\right] ~.  \label{energy d}
\end{equation}%
The left hand side vanishes for arbitrary choices of $\dot{\rho}$ provided%
\begin{equation}
\eta \dot{\phi}+\frac{\eta ^{2}}{2m}(\partial _{a}\phi )^{2}+V-\frac{\mu
\eta ^{2}}{2m^{2}}\frac{\nabla ^{2}\rho ^{1/2}}{\rho ^{1/2}}=0~.  \label{SEb}
\end{equation}

Equations (\ref{SEa}) and (\ref{SEb}) are the coupled dynamical equations we
seek. They describe entropic diffusion and energy conservation. The
evolution of $\rho (x,t)$, eq.(\ref{SEa}), is guided by $\phi (x,t)$; the
evolution of $\phi (x,t)$, eq.(\ref{SEb}), is determined by $\rho (x,t)$.
The evolving geometry of the manifold $\mathcal{M}$ enters through $\phi
(x,t)$.

\subsection{Classical limits}

Before proceeding further we note that writing $S_{HJ}=\eta \phi $ in
equations (\ref{curr osmo}) and (\ref{SEb}) and taking the limit $\eta
\rightarrow 0$ with $S_{HJ}$, $m$, and $\mu $ fixed leads to 
\begin{equation}
mv_{a}=\,\partial _{a}S_{HJ}\quad \text{and\quad }u_{a}=0~,
\end{equation}%
and to the Hamilton-Jacobi equation%
\begin{equation}
\dot{S}_{HJ}+\frac{1}{2m}(\partial _{a}S_{HJ})^{2}+V=0~.  \label{HJ}
\end{equation}%
This suggests that the constant $m$ be interpreted as the inertial mass.
Furthermore, eq.(\ref{drift future}) tells us that the particle is expected
to move along the entropy gradient, while eq.(\ref{fluc}), 
\begin{equation}
\left\langle \Delta w^{a}\right\rangle =0\quad \text{and}\quad \left\langle
\Delta w^{a}\Delta w^{b}\right\rangle =\frac{\eta }{m}\Delta t\,\delta
^{ab}\rightarrow 0~,  \label{Zero Fluct}
\end{equation}%
says that the fluctuations about the expected trajectory vanish. We conclude
that in the limit $\eta \rightarrow 0$ entropic dynamics reproduces
classical mechanics with classical trajectories following the entropy
gradient. A similar classical limit can also be attained for fixed $\eta $
provided the mass $m$ is sufficiently large.

The limit $\mu \rightarrow 0$ for fixed $\eta $, $S_{HJ}$, and $m$ is also
interesting. This situation is also ruled by the classical Hamilton-Jacobi
equation (\ref{HJ}), but the osmotic velocity does not vanish, 
\begin{equation}
mv_{a}=\,\partial _{a}S_{HJ}\quad \text{and\quad }mu_{a}=\eta \partial
_{a}\log \rho ^{1/2}~~.
\end{equation}%
The expected trajectory also lies along a classical path but now, however,
it does not coincide with the entropy gradient. More important perhaps is
the fact that the fluctuations $\Delta w^{a}$ about the classical trajectory
do not vanish. The limit $\mu \rightarrow 0$ is a different
\textquotedblleft classical\textquotedblright\ limit; whether it corresponds
to an actual physical situation remains to be seen.

\subsection{The Schr\"{o}dinger equation}

Next we show that, with one very interesting twist, the dynamical equations (%
\ref{SEa}) and (\ref{SEb}) turn out to be equivalent to the Schr\"{o}dinger
equation. We can always combine the functions $\rho $ and $\phi $ into a
complex function 
\begin{equation}
\Psi =\rho ^{1/2}\exp (i\phi )~.
\end{equation}%
Computing its time derivative, 
\begin{equation}
\dot{\Psi}=\left( \frac{\dot{\rho}}{2\rho }+i\dot{\phi}\right) \Psi ~,
\label{Psi dot}
\end{equation}%
and using eqs. (\ref{SEa}) and (\ref{SEb}) leads to 
\begin{equation}
i\eta \dot{\Psi}=-\frac{\eta ^{2}}{2m}\nabla ^{2}\Psi +V\Psi +\frac{\eta ^{2}%
}{2m}\left( 1-\frac{\mu }{m}\right) \frac{\nabla ^{2}(\Psi \Psi ^{\ast
})^{1/2}}{(\Psi \Psi ^{\ast })^{1/2}}\Psi ~.  \label{SEc}
\end{equation}%
This reproduces the Schr\"{o}dinger equation, 
\begin{equation}
i\hbar \frac{\partial \Psi }{\partial t}=-\frac{\hbar ^{2}}{2m}\nabla
^{2}\Psi +V\Psi ~,  \label{SE}
\end{equation}%
provided the current and osmotic masses are equal, $m=\mu $, and $\eta $ is
identified with Planck's constant, $\eta =\hbar $.

But why should the osmotic mass be precisely equal to the inertial mass? Why
can't we say that entropic dynamics predicts a non-linear generalization of
quantum theory? This question is so central to quantum theory that we devote
the next section to it. But before that we note that the non-linearity is
undesirable both for experimental and theoretical reasons. On one hand,
various types of non-linearities have been ruled out experimentally to an
extreme degree through precision experiments on the Lamb shift \cite{Smolin
86a} and even more so in hyperfine transitions \cite{Bollinger 89}. On the
other hand, from the theory side it is the fact that time evolution
preserves linear superpositions that leads to the superposition principle
and makes Hilbert spaces useful. In addition, there is a consistency
argument that links the linearity of the Hilbert space and the linearity of
time evolution \cite{Caticha 98}. Retaining one and not the other leads to
inconsistently assigned amplitudes showing that the very concept of quantum
amplitudes is a reflection of linearity. And, as if that were not enough, it
has also been shown that in the presence of non-linear terms entangled
particles could be used to achieve\ superluminal communication \cite{Gisin
90}. Therefore it is extremely probable that the identity of inertial and
osmotic mass is exact.

There is another mystery in quantum theory---the central role played by
complex numbers---which turns out to be related to these issues. The
dynamical equations (\ref{SEa}) and (\ref{SEb}) contain no complex numbers.
It is true that they contain two degrees of freedom $\rho $ and $\phi $ and
that these two can be combined into a single complex number $\Psi =\rho
^{1/2}e^{i\phi }$. This is a triviality, not a mystery: the dynamical
equations can always be reformulated into an equation for $\Psi $ \emph{and
its conjugate }$\Psi ^{\ast }$. The statement that complex numbers play a
fundamental role in quantum theory is the non-trivial assertion that the
equation of evolution contains \emph{only }$\Psi $\emph{\ }and not $\Psi $
and also its conjugate $\Psi ^{\ast }$. In the entropic approach both the
linear time evolution and the special role of complex numbers are linked
through the equality $m=\mu $.

\section{A quantum equivalence principle}

The generalization to $N$ particles is straightforward. As indicated at the
end of section 2, the configuration space has $3N$ dimensions and the system
is represented by a point $x=x^{A}=(x^{a_{1}},x^{a_{2}}\ldots )$. The
corresponding Fokker-Planck equation [see eqs.(\ref{gamma AB}), (\ref{FP b})
and (\ref{curr})] is 
\begin{equation}
\partial _{t}\rho =-\frac{1}{\tau }\gamma ^{AB}\partial _{A}(\rho \partial
_{B}\phi )=-\tsum\limits_{n=1}^{N}\partial _{a_{n}}\left( \rho
v_{n}^{a_{n}}\right)  \label{FP c}
\end{equation}%
where $\phi (x,t)$ is given by eq.(\ref{phase}). The current and osmotic
velocities are%
\begin{equation}
v_{n}^{a_{n}}=\frac{\sigma _{n}^{2}}{\tau }\,\partial ^{a_{n}}\phi \quad 
\text{and}\quad \mu _{n}^{a_{n}}=-\frac{\sigma _{n}^{2}}{\tau }\,\partial
^{a_{n}}\log \rho ^{1/2}~,
\end{equation}%
and the conserved energy is

\begin{equation}
E=\int d^{3N}x\,\rho (x,t)\left( A\gamma _{AB}v^{A}v^{B}+B\gamma
_{AB}u^{A}u^{B}+V(x)\right) \,.  \label{energy e}
\end{equation}%
Introducing the inertial (or current) and osmotic masses,%
\begin{equation}
m_{n}=\frac{2A}{\sigma _{n}^{2}}\quad \text{and\quad }\mu _{n}=\frac{2B}{%
\sigma _{n}^{2}}~,  \label{masses}
\end{equation}%
and the constant $\eta =2A/\tau $, eqs.(\ref{FP c}) and (\ref{energy e})
become 
\begin{equation}
\partial _{t}\rho =-\tsum\limits_{n}\frac{\eta }{m_{n}}\partial
_{a_{n}}\left( \rho \partial ^{a_{n}}\phi \right) ~,  \label{FP d}
\end{equation}%
\begin{equation}
E[\rho ,S]=\int d^{3N}x\,\rho \left( \tsum\limits_{n}[\frac{\eta ^{2}}{2m_{n}%
}(\partial _{a_{n}}\phi )^{2}+\frac{\mu _{n}\eta ^{2}}{8m_{n}^{2}}(\partial
_{a_{n}}\log \rho )^{2}]+V(x)\right) ~.  \label{energy f}
\end{equation}%
Imposing, as before, that $\dot{E}-\tint \rho \dot{V}=0$ for arbitrary
choices of $\dot{\rho}$ leads to the modified Hamilton-Jacobi equation, 
\begin{equation}
\eta \dot{\phi}+\tsum\limits_{n}[\frac{\eta ^{2}}{2m_{n}}(\partial
_{a_{n}}\phi )^{2}-\frac{\mu _{n}\eta ^{2}}{2m_{n}^{2}}\frac{\nabla
_{n}^{2}\rho ^{1/2}}{\rho ^{1/2}}]+V=0~.  \label{HJ a}
\end{equation}%
Finally, the two eqs.(\ref{FP d}) and (\ref{HJ a}) can be combined into a
single equation for the complex wave function, $\Psi =\rho ^{1/2}e^{i\phi }$%
, 
\begin{equation}
i\eta \dot{\Psi}=\tsum\limits_{n}\frac{-\eta ^{2}}{2m_{n}}[\nabla
_{n}^{2}-\left( 1-\frac{\mu _{n}}{m_{n}}\right) \frac{\nabla _{n}^{2}(\Psi
\Psi ^{\ast })^{1/2}}{(\Psi \Psi ^{\ast })^{1/2}}]\Psi +V\Psi ~.  \label{SEe}
\end{equation}

Eq.(\ref{masses}) shows that the ratio of osmotic to inertial mass turns out
to be a universal constant, the same for all particles: $\mu _{n}/m_{n}=B/A$%
. This can be traced to a choice of energy that reflects the translational
and rotational symmetries of the configuration space. But why should $\mu
_{n}=m_{n}$ \emph{exactly}? To see this we go back to eq.(\ref{energy f}).
We can always change units and rescale $\eta $ and $\tau $ by some constant $%
\kappa $ into $\eta =\kappa \eta ^{\prime }$, $\tau =\tau ^{\prime }/\kappa $%
. If we also rescale $\phi $ into $\phi =\phi ^{\prime }/\kappa $, then eqs.(%
\ref{FP d}) and (\ref{energy f}) become 
\begin{equation}
\partial _{t}\rho =-\tsum\limits_{n}\frac{\eta ^{\prime }}{m_{n}}\partial
_{a_{n}}\left( \rho \partial ^{a_{n}}\phi ^{\prime }\right) ~,
\end{equation}%
\begin{equation}
E[\rho ,S]=\int d^{3N}x\,\rho \left( \tsum\limits_{n}[\frac{\eta ^{\prime 2}%
}{2m_{n}}(\partial _{a_{n}}\phi ^{\prime })^{2}+\frac{\mu _{n}\kappa
^{2}\eta ^{\prime 2}}{8m_{n}^{2}}(\partial _{a_{n}}\log \rho )^{2}]+V\right)
~.
\end{equation}%
Following the same steps that led to eq.(\ref{SEd}), we can introduce a 
\emph{different }wave function $\Psi ^{\prime }=\rho ^{1/2}\exp (i\phi
^{\prime })$ which satisfies 
\begin{equation}
i\eta ^{\prime }\dot{\Psi}^{\prime }=\tsum\limits_{n}\frac{-\eta ^{\prime 2}%
}{2m_{n}}[\nabla _{n}^{2}-\left( 1-\frac{\mu _{n}\kappa ^{2}}{m_{n}}\right) 
\frac{\nabla _{n}^{2}(\Psi ^{\prime }\Psi ^{\prime \ast })^{1/2}}{(\Psi
^{\prime }\Psi ^{\prime \ast })^{1/2}}]\Psi ^{\prime }+V\Psi ^{\prime }~.
\end{equation}%
Since the mere rescaling by $\kappa $ can have no physical implications the
different \textquotedblleft regraduated\textquotedblright\ theories are all
equivalent and it is only natural to use the simplest one: we choose $\kappa
=(A/B)^{1/2}$ so that $\mu _{n}\kappa ^{2}=m_{n}$ and we can rescale the old 
$\mu _{n}$ to a new osmotic mass $\mu _{n}^{\prime }=\mu _{n}\kappa
^{2}=m_{n}$.

The net result is that the non-linear terms drop out. Dropping the prime on $%
\Psi ^{\prime }$ and identifying the rescaled value $\eta ^{\prime }$ with
Planck's constant $\hbar $, leads to the linear Schr\"{o}dinger equation, 
\begin{equation}
i\hbar \dot{\Psi}=\tsum\limits_{n}\frac{-\hbar ^{2}}{2m_{n}}\nabla
_{n}^{2}\Psi +V\Psi ~.
\end{equation}

We conclude that \emph{for} \emph{any positive value of the original
coefficients }$\mu _{n}$\emph{\ it is always possible to regraduate }$\eta $%
\emph{, }$\phi $\emph{\ and }$\mu _{n}$\emph{\ to a physically equivalent
but more convenient description where the Schr\"{o}dinger equation is linear
and complex numbers attain a special significance}. From this entropic
perspective the linear superposition principle and the complex Hilbert
spaces are important because they are convenient, but not because they are
fundamental---a theme that was also explored in \cite{Caticha 98}.

These considerations remind us of Einstein's original argument for the
equivalence principle: We accept the complete physical equivalence of a
gravitational field with the corresponding acceleration of the reference
frame because this offers a natural explanation of the equality of inertial
and gravitational masses and opens the door to an explanation of gravity in
purely geometrical terms.

Similarly, in the quantum case \emph{we accept the complete equivalence of
quantum and statistical fluctuations because this offers a natural
explanation of the Schr\"{o}dinger equation---its linearity, its unitarity,
the role of complex numbers, the equality of inertial and osmotic masses.
Furthermore, it opens the door to explaining quantum theory as an example of
statistical inference---entropic dynamics on a suitably non-trivial evolving
manifold.}

\section{Entropic time \emph{vs.} \textquotedblleft
physical\textquotedblright\ time}

Now that the dynamics has been more fully developed we should revisit the
question of time. Entropic time has turned out to be useful in ordering the
inferential sequence of small changes but it is not at all clear that this
order has anything to do with the order relative to a presumably more
fundamental \textquotedblleft physical\textquotedblright\ time. If so, why
does `entropic time' deserve to be called `time' at all?

The answer is that the systems we are typically concerned with include, in
addition to the particles of interest, also another system that one might
call the \textquotedblleft clock\textquotedblright . The goal is to make
inferences about correlations among the particles themselves and with the
various states of the clock. Whether the inferred sequence of states of the
particle-clock composite agrees with the order in \textquotedblleft
physical\textquotedblright\ time or not turns out to be quite irrelevant. It
is only the correlations among the particles and the clock that are
observable and not their \textquotedblleft absolute\textquotedblright\ order.

This is an idea that demands a more explicit discussion. Here we show how it
gives rise to the notion of simultaneity that turned out to be central to
our definition of an instant in section 4.1.

Consider a single particle. From the probability of a single step, eq.(\ref%
{Prob xp/x}) or (\ref{Prob xp/x b}), we can calculate the probability of any
given sequence of (short) steps $\{x,x_{1},\ldots ,x_{n},\ldots \}$. Since
the path is an ordered sequence of events when two events lie on the same
path it is meaningful to assert that one is earlier (in the entropic time
sense) than the other:\ $x_{n}$ is earlier than $x_{n+1}$. The actual path,
however, is uncertain: how do we compare possible events along different
paths? We need a criterion that will allow us to decide whether an event $%
x^{\prime }$ reached along one path is earlier or later than another event $%
x^{\prime \prime }$ reached along a different path. This is where the clock
comes in. The role of the clock can be played, for example, by a
sufficiently massive particle. This guarantees that the clock follows a
deterministic classical trajectory $x_{C}=\bar{x}_{C}(t)$ given by eqs.(\ref%
{HJ}) and (\ref{Zero Fluct}) and that it remains largely unaffected by the
motion of the particle.

The idea is that when we compute the probability that, say, after $n$ steps
the particle is found at the point $x_{n}$ we implicitly \emph{assume} that
its three coordinates $x_{n}^{1}$, $x_{n}^{2}$, and $x_{n}^{3}$ are attained 
\emph{simultaneously}. This is part of our \emph{definition} of an instant.
We make the same \emph{definition} for composite systems. In particular, for
the particle-clock system, $x_{n}^{A}=(x_{n}^{a},x_{Cn}^{\alpha })$, the
coordinates of the particle $x_{n}^{a}$ ($a=1,2,3$) are taken to be
simultaneous with the remaining coordinates that describe the clock $%
x_{Cn}^{\alpha }$ ($\alpha =4,5,\ldots $). Thus, when we say that at the $n$%
th step the particle is at $x_{n}^{a}$ while the clock is at $x_{Cn}^{\alpha
}$ it is implicit that these positions are attained \emph{at the same time}.

By \textquotedblleft the time is $t$\textquotedblright\ we will just mean
that \textquotedblleft the clock is in its state $x_{C}=\bar{x}_{C}(t)$%
.\textquotedblright\ We say that the possible event that the particle
reached $x^{\prime }$ along one path is simultaneous with another possible
event $x^{\prime \prime }$ reached along a different path when both are
simultaneous with the same state $\bar{x}_{C}(t)$ of the clock: then we say
that $x^{\prime }$ and $x^{\prime \prime }$ happen \textquotedblleft at the
same time $t$.\textquotedblright\ This justifies using the distribution $%
\rho (x,t)$ as the definition of an instant of time.

In the end the justification for the assumptions underlying entropic
dynamics lies in experiment. The ordering scheme provided by entropic time
allows one to predict correlations. Since these predictions, which are given
by the Schr\"{o}dinger equation, turn out to be empirically successful one
concludes that nothing deeper or more \textquotedblleft
physical\textquotedblright\ than entropic time is needed. A similar claim
has been made by J. Barbour in his relational approach to time in the
context of classical dynamics \cite{Barbour 94}.

\section{Dynamics in an external electromagnetic field}

Entropic dynamics is derived from the minimal assumptions that the extra
variables $y$ are intrinsically uncertain and that motion consists of a
succession of short steps. These two pieces of information are taken into
account through the two constraints (\ref{constraint p}) and (\ref{short
step}). Special circumstances may however require additional constraints.

\subsection{An additional constraint}

Consider a single particle placed in an external field the action of which
is to constrain the expected component of displacements along a certain
direction represented by the unit covector $n_{a}(x)$. This effect is
represented by the constraint 
\begin{equation}
\langle \Delta x^{a}n_{a}(x)\rangle =C(x)~,
\end{equation}%
where the spatial dependence of $C(x)$ reflects the non-uniform intensity of
the external field. It is convenient to define the magnitude of the external
field in terms of the effect it induces. Thus we introduce the external
field 
\begin{equation}
A_{a}(x)\propto \frac{n_{a}(x)}{C(x)}~
\end{equation}%
and the constraint is 
\begin{equation}
\langle \Delta x^{a}A_{a}(x)\rangle =C~,  \label{constraint A}
\end{equation}%
where $C$ is some constant that reflects the strength of the coupling to $%
A_{a}$.

\subsection{Entropic dynamics}

The transition probability $P(x^{\prime }|x)$ is that which maximizes the
entropy $\mathcal{S}[P,Q]$ in (\ref{Sppi b}) subject to the old constraints
plus the new constraint (\ref{constraint A}). The result is%
\begin{equation}
P(x^{\prime }|x)=\frac{1}{\zeta (x,\alpha ,\beta )}e^{S(x^{\prime })-\frac{1%
}{2}\alpha \Delta \ell ^{2}(x^{\prime },x)-\beta \Delta x^{a}A_{a}(x)}~,
\label{Prob xp/x c}
\end{equation}%
where 
\begin{equation}
\zeta (x,\alpha ,\beta )=\int dx^{\prime }\,e^{S(x^{\prime })-\frac{1}{2}%
\alpha \Delta \ell ^{2}(x^{\prime },x)-\beta \Delta x^{a}A_{a}(x)}~,
\end{equation}%
and the Lagrange multiplier $\beta $ is determined from the constraint eq.(%
\ref{constraint A}), 
\begin{equation}
\frac{\partial }{\partial \beta }\log \zeta (x,\alpha ,\beta )=-C~.
\end{equation}

From here on the argument follows closely the previous sections. For large $%
\alpha $ the transition probability (\ref{Prob xp/x c}) can be written as 
\begin{equation}
P(x^{\prime }|x)\propto \exp \left[ -\frac{m}{2\hbar \Delta t}\delta
_{ab}\left( \Delta x^{a}-\Delta \bar{x}^{a}\right) \left( \Delta
x^{b}-\Delta \bar{x}^{b}\right) \right] ~,  \label{Prob xp/x d}
\end{equation}%
where we used (\ref{alpha}), (\ref{eta}), and units have been regraduated to
set $\eta =\hbar $. Therefore, the displacement $\Delta x^{a}$ can be
expressed in terms of a expected drift plus a fluctuation, $\Delta
x^{a}=\Delta \bar{x}^{a}+\Delta w^{a}$, where%
\begin{equation}
\left\langle \Delta x^{a}\right\rangle =\Delta \bar{x}^{a}=b^{a}\Delta
t\quad \text{where}\quad b^{a}=\frac{\hbar }{m}\delta ^{ab}[\partial
_{b}S-\beta A_{b}]~,  \label{drift future EM}
\end{equation}%
\begin{equation}
\left\langle \Delta w^{a}\right\rangle =0\quad \text{and}\quad \left\langle
\Delta w^{a}\Delta w^{b}\right\rangle =\frac{\hbar }{m}\Delta t\,\delta
^{ab}~.
\end{equation}%
Once again, for short steps the dynamics is dominated by the fluctuations.
The only difference is the replacement of $\partial S$ by the gauge
invariant combination $\partial S-\beta A$. Small changes accumulate
according to the FP equation (\ref{FP b}) but now the current velocity is no
longer given by eq.(\ref{curr}) but rather by%
\begin{equation}
v^{a}=\frac{\hbar }{m}(\,\partial ^{a}\phi -\beta A^{a})~,  \label{curr EM}
\end{equation}%
and the FP equation is 
\begin{equation}
\dot{\rho}=-\partial _{a}\left( \rho v^{a}\right) =-\frac{\hbar }{m}\partial
^{a}[\rho (\partial _{a}\phi -\beta A_{a})]~,  \label{FP EM}
\end{equation}%
$\phi $ is still given by (\ref{phase}) and the osmotic velocity (\ref{osmo}%
) remains unchanged.

The energy functional is the same as (\ref{energy b}), but now $v$ is given
by eq.(\ref{curr EM}), 
\begin{equation}
E=\int dx\,\rho \left( \frac{\hbar ^{2}}{2m}(\partial _{a}\phi -\beta
A_{a})^{2}+\frac{\hbar ^{2}}{8m}(\partial _{a}\log \rho )^{2}+V\right) ~,
\label{energy EM}
\end{equation}%
where we set $\mu =m$ and $\eta =\hbar $.

It is simplest to start with static external potentials, $\dot{V}=0$ and $%
\dot{A}=0$, so that the energy is conserved, $\dot{E}=0$. Just as before
after taking the time derivative, integrating by parts, and imposing that $%
\dot{E}=0$ for arbitrary choices of $\dot{\rho}$, we get 
\begin{equation}
\hbar \dot{\phi}+\frac{\hbar ^{2}}{2m}(\partial _{a}\phi -\beta A_{a})^{2}+V-%
\frac{\hbar ^{2}}{2m}\frac{\nabla ^{2}\rho ^{1/2}}{\rho ^{1/2}}=0~.
\label{HJ EM}
\end{equation}%
Equations (\ref{FP EM}) and (\ref{HJ EM}) are the coupled equations for $%
\rho $ and $\phi $ that describe entropic dynamics in the external potential 
$A_{a}$.

Setting $S_{HJ}=\eta \phi $ and taking the classical limit $\hbar
\rightarrow 0$ leads to the classical Hamilton-Jacobi equation in an
external electromagnetic field showing that the Lagrange multiplier $\beta $
plays the role of electric charge. More precisely,%
\begin{equation}
\beta =\frac{e}{\hbar c}~,
\end{equation}%
where $e$ is the electric charge and $c$ is the speed of light. Thus, in
entropic dynamics electric charge is a Lagrange multiplier that regulates
the response to the external electromagnetic potential $A_{a}$. (If desired
we can further separate $V$ into electric and non-electric components, $%
V=eA_{0}+V^{\prime }$, but this is not needed for our present purposes.)

As before, the Schr\"{o}dinger equation results from combining the functions 
$\rho $ and $\phi $ into the wave function, $\Psi =\rho ^{1/2}\exp (i\phi )$%
. Computing the time derivative $\dot{\Psi}$ using eqs.(\ref{FP EM}) and (%
\ref{HJ EM}) leads to the Schr\"{o}dinger equation, 
\begin{equation}
i\hbar \frac{\partial \Psi }{\partial t}=\frac{\hbar ^{2}}{2m}(i\partial
_{a}-\frac{e}{\hbar c}A_{a})^{2}\Psi +V\Psi ~,  \label{SEd}
\end{equation}

The derivation above assumed that energy is conserved, $\dot{E}=0$, which is
true when the external potentials are static, $\dot{V}=0$ and $\dot{A}=0$,
but this limitation is easily lifted. For time-dependent potentials the
relevant energy condition must take into account the work done by external
sources: we require that the energy increase at the rate 
\begin{equation}
\dot{E}=\int dx\,\rho (\dot{V}+\frac{e}{c}\rho v^{a}\dot{A}_{a})~.~
\end{equation}%
The net result is that equations (\ref{HJ EM}) and (\ref{SEd}) remain valid
for time-dependent external potentials.

\subsection{Gauge invariance}

We have seen that in entropic dynamics the phase of the wave function
receives a statistical interpretation, $\phi =S-\log \rho ^{1/2}$. On the
other hand, without any physical consequences, the phase can be shifted by
an arbitrary amount, 
\begin{equation}
\phi (x,t)\rightarrow \phi ^{\prime }(x,t)=\phi (x,t)+\beta \chi (x,t)~,
\end{equation}%
provided the potential is transformed appropriately, $A_{a}\rightarrow
A_{a}^{\prime }=A_{a}+\partial _{a}\chi $. This raises several questions.

First, how is the statistical interpretation of $\phi $ affected by the
possibility of gauge transformations? The straighforward answer is that $%
\phi $ reflects a combination of several effects---the extra variables
(through their entropy $S$), the osmotic effect of diffusion (through the
density $\rho $), and the choice of potential (through the function $\chi $%
)---but these separate contributions are not necessarily easy to
disentangle. Indeed, eq.(\ref{drift future EM}) for the drift velocity shows
that the dynamics depends on $S$ and on $A$ only through the combination $%
\partial S-\beta A$. Therefore we can envision two situations that are
informationally inequivalent: one is characterized by entropy $S$ and
constraint $\langle \Delta x^{a}A_{a}\rangle =C$, the other by a different
entropy $S^{\prime }$ and also by a different constraint $\langle \Delta
x^{a}A_{a}^{\prime }\rangle =C$. Remarkably they lead to exactly the same
physical predictions provided the entropies and potentials are related by $%
S^{\prime }=S+\beta \chi $ and $A^{\prime }=A+\partial \chi $ where $\chi
(x,t)$ is some arbitrary function. Thus local phase invariance can be
interpreted as \emph{local entropy invariance}.

A second question was first raised in the context of stochastic mechanics
and concerns the single- or multi-valuedness of phases and wave functions.
Wallstrom \cite{Wallstrom 94} noted that when stochastic mechanics is
formulated \emph{\`{a} la} Nelson \cite{Nelson 66} the current velocity $%
\vec{v}$ is postulated to be the gradient of some locally defined function $%
\phi $. Now, being a local gradient does not imply that $\vec{v}$ will also
be a global gradient and therefore both the phases $\phi $ and their
corresponding wave functions $\Psi $ will, in general, be
multi-valued---which is unsatisfactory. A possible way out is to formulate
stochastic mechanics in terms of an action principle \cite{Guerra Morato 83}%
. Then the current velocity is indeed a global gradient and both phases and
wave functions are single-valued. But this is a problem too: single-valued
phases can be too restrictive and exclude physically relevant states. For
example, the usual way to describe states with non-zero angular momentum is
to use multi-valued phases (the azimuthal angle) while requiring that the
corresponding wave functions remain single-valued.

The same questions can be raised in entropic dynamics and also within the
standard quantum framework. Why should wave functions be single-valued? The
answer we favor is essentially the same offered by Pauli in the context of
standard quantum mechanics \cite{Pauli 39}. He suggested that the criterion
for admissibility for wave functions is that they must form a basis for a
representation of the transformation group (\emph{e.g.}, the rotation group)
that happens to be pertinent to the problem at hand. Pauli's criterion is
extremely natural from the perspective of a theory of inference: in any
physical situation symmetries constitute the most common and most obviously
relevant pieces of information.

In entropic dynamics the entropy $S(x,t)$ and the probability density $\rho
(x,t)$ are single-valued functions. Therefore, a natural choice is that the
phase, $\phi =S-\log \rho ^{1/2}$, be single-valued too. A situation with
non-vanishing angular momentum can be handled through a constraint. For
example, one can use a single-valued phase and an appropriately chosen
vector potential---which might perhaps be a pure gauge, $A_{a}=-\partial
_{a}\chi $. Alternatively, we can gauge the potential away to $A_{a}^{\prime
}=0$ and use a multi-valued phase, $\phi ^{\prime }=S-\log \rho ^{1/2}+\beta
\chi $. Which of these two options is to be preferred depends on whether the
goal is clarity of interpretation or simpler mathematics. As for the
appropriate choice of potential, $A_{a}=\partial _{a}\chi $, we adopt
Pauli's criterion: the admissible wavefunctions---that is, the various
functions $(\rho ,S,\chi )$ that appear in the formalism---must form a basis
for a representation of the pertinent transformation group.

\section{Summary and Conclusions}

Our goal has been to derive quantum theory as an example of entropic
inference. The challenge is to develop a framework that clarifies the
conceptual difficulties that have plagued quantum theory since its inception
while still reproducing its undeniable experimental successes. This means
that to the extent that what has been derived is quantum mechanics and not
some other theory we should not expect predictions that deviate from those
of the standard quantum theory---at least not in the non-relativistic regime
discussed in this paper. On the other hand, the motivation behind this whole
program lies in the conviction that it is the clarification and removal of
conceptual difficulties that will eventually allow us to extend physics to
other realms---gravity, cosmology---where the status of quantum theory is
more questionable.

The framework of entropic inference is of general applicability. Its
application to any particular problem requires assumptions that specify the
intended subject matter and those pieces of information are considered
relevant. The main assumptions can be summarized as follows:

\noindent \textbf{(a)} The goal is to predict the positions $x$ of some
point particles. Since the information available is limited we can at best
obtain a probability distribution $\rho (x)$ in the configuration space $%
\mathcal{X}$. We assume that $\mathcal{X}$ is flat, and that it is isotropic
or anisotropic depending on whether the particles are identical or not.

\noindent \textbf{(b)} We assume that the world includes other things in
addition to the particles: these extra things are described by variables $y$
that can influence and in turn can be influenced by the particles. The
uncertainty in the values of $y$ is described by distributions $p(y|x)$ in a
statistical manifold $\mathcal{M}$. The theory is robust in the sense that
its predictions are insensitive to most details about the extra variables.

\noindent \textbf{(c)} We assume that large changes result from the
accumulation of many successive short steps. The transition probability for
a short step $P(x^{\prime }|x)$ is found using the method of maximum
entropy. This requires assumptions about the prior (which we take to be
uniform) and constraints (that changes happen continuously and that after
each short step the new $p(y^{\prime }|x^{\prime })$ remains within the same
statistical manifold $\mathcal{M}$). The result is that the dynamics of the
particles is driven by the entropy $S(x)$ of the extra variables.

\noindent \textbf{(d)} A notion of time is introduced in order to keep track
of the accumulation of small changes. This requires assumptions about what
constitutes an `instant' and about how time is `constructed' as a succession
of such instants. The choice of interval between instants is a matter of
convenience---we choose a notion of duration that reflects the translational
symmetry of the configuration space. The result is that the distribution $%
\rho $ evolves according to a Fokker-Planck equation.

\noindent \textbf{(e)} We assume that the particles react back and affect
the entropy $S(x)$ of the extra variables in such a way that there is a
conserved `energy' $E[\rho ,S]=\func{const}$. The specifics of this
interaction are described through the functional form of $E[\rho ,S]$.

\noindent \textbf{(f)} Electromagnetic interactions are described by
including an additional constraint on the expected displacement along a
certain field $A_{a}(x)$.

No further assumptions are made.

The statistical model is specified by several parameters, $\{\sigma
_{n}^{2},\tau ,A,B,\beta \}$. The anisotropy of configuration space for
non-identical particles is parametrized by $\sigma _{n}^{2}$ with $n=1\ldots
N$; $\tau $ defines units of time; $A$ and $B$ parametrize the relative
strengths of the current and osmotic terms in the energy functional; and,
finally, $\beta $ is the Lagrange multiplier associated to the
electromagnetic constraint. These parameters can be suitably regraduated and
combined with each other into the familiar set which includes the masses and
charges of the particles and Planck's constant.

We conclude with a summary of our conclusions.

\noindent \textbf{On epistemology \emph{vs.} ontology:} Quantum theory has
been derived as an example of entropic dynamics. In this model
\textquotedblleft reality\textquotedblright\ is reflected in the positions
of the particles and the values of the extra variables, and our
\textquotedblleft limited information about reality\textquotedblright\ is
represented in the probabilities as they are updated to reflect the
physically relevant\ constraints.

\noindent \textbf{Quantum non-locality:} Entropic dynamics may appear
classical because no \textquotedblleft quantum\textquotedblright\
probabilities were introduced. But this is deceptive. Probabilities, in this
approach, are neither classical nor quantum; they are merely tools for
inference. Phenomena that would normally be considered non-classical, such
as non-local correlations, emerge naturally from constraints in
configuration space which include the osmotic or, equivalently, the quantum
potential terms in the energy functional.

The presence of a quantum potential may suggest a connection between our
(epistemological) entropic dynamics and Bohm's ontological interpretation 
\cite{Bohm Hiley 93}. There is none. It is true that both theories agree on
the same Schr\"{o}dinger equation, and therefore on the same modified
version of the Hamilton-Jacobi equation. But Bohmian mechanics is meant to
reflect \textquotedblleft reality\textquotedblright ; particles are supposed
to follow smooth causal trajectories along the gradient of the phase, $%
\nabla \phi $. In contrast, entropic dynamics reflects information;
particles follow non-differentiable trajectories and it is the probability
distribution $\rho $, not the particles, that evolves along $\nabla \phi $.

\noindent \textbf{On interpretation:} Ever since Born the magnitude of the
wave function $|\Psi |^{2}=\rho $ has received a statistical interpretation.
Within the entropic dynamics approach the phase of the wave function is also
recognized as a feature of purely statistical origin. When electromagnetic
interactions are introduced the gauge invariance is interpreted as an
invariance under local entropy transformations.

\noindent \textbf{On dynamical laws:} The principles of entropic inference
form the backbone of this approach to dynamics. The requirement that an
energy be conserved is an important piece of information (\emph{i.e.}, a
constraint) which will probably receive its full justification once a
completely relativistic version of entropic dynamics is developed.

\noindent \textbf{On time:} The derivation of laws of physics as examples of
inference requires an account of the concept of time. Entropic time is
modelled as an ordered sequence of instants with the natural measure of
duration chosen to simplify the description of motion. We argued that
whether the entropic order agrees with an objective order in an external
\textquotedblleft physical\textquotedblright\ time turns out to be an
empirically inaccessible question, and in this sense, the notion of a
\textquotedblleft physical\textquotedblright\ time is not needed. Most
interestingly, the entropic model of time explains the arrow of time.

\noindent \textbf{Equivalence principle:} The derivation of the Schr\"{o}%
dinger equation from entropic inference led to an interesting analogy with
general relativity. The statistical manifold $\mathcal{M}$ is not a fixed
background but actively participates in the dynamics.

\noindent \textbf{Acknowledgments:} I would like to thank C. Rodr\'{\i}guez
and N. Caticha for their insights into entropy and inference, and D.
Bartolomeo for pointing out an important mistake in an earlier version of
this paper. My deep appreciation also goes to C. Cafaro, A. Giffin, P.
Goyal, D. Johnson, K. Knuth, S. Nawaz, M. Reginatto, and C.-Y. Tseng for
many discussions. This research was supported by the University at
Albany-SUNY while on sabbatical leave. The hospitality of the Perimeter
Institute for Theoretical Physics (Waterloo, Canada) through an extended
visit is also gratefully acknowledged.

\section*{Appendix}

We claim that the distribution $P(x^{\prime },y^{\prime }|x)$ given by (\ref%
{Prob xp/x}) and (\ref{constraint p}) yields the absolute global maximum of
the relative entropy $\mathcal{S}[P,Q]$ subject to the constraint (\ref%
{short step}). Substituting (\ref{Prob xp/x}) into (\ref{Sppi b}) the
conjectured maximum value is 
\begin{equation}
\mathcal{S}[P,Q]=\frac{1}{2}\alpha \langle \Delta \ell ^{2}\rangle -\log
(\zeta \gamma ^{1/2})=\mathcal{S}_{\max }~.
\end{equation}%
On the other hand, the entropy $\mathcal{K}[\tilde{P},P]$ of any
distribution $\tilde{P}(x^{\prime },y^{\prime }|x)$ relative to $P(x^{\prime
},y^{\prime }|x)$ satisfies the Gibbs inequality \cite{Jaynes 03}, 
\begin{equation}
\mathcal{K}[\tilde{P},P]=-\int dx^{\prime }dy^{\prime }\,\tilde{P}(x^{\prime
},y^{\prime }|x)\log \frac{\tilde{P}(x^{\prime },y^{\prime }|x)}{P(x^{\prime
},y^{\prime }|x)}\leq 0~,
\end{equation}%
with equality if and only if $\tilde{P}=P$. Introducing the prior $Q$ and
rearranging we get 
\begin{equation}
\mathcal{S}[\tilde{P},Q]\leq -\int dx^{\prime }dy^{\prime }\,\tilde{P}%
(x^{\prime },y^{\prime }|x)\log \frac{P(x^{\prime },y^{\prime }|x)}{%
Q(x^{\prime },y^{\prime }|x)}~.
\end{equation}%
We require that $\tilde{P}$ satisfy the same constraints (\ref{constraint p}%
) and (\ref{short step}) and use (\ref{Prob xp/x}) to get 
\begin{eqnarray}
\mathcal{S}[\tilde{P},Q] &\leq &-\int dx^{\prime }dy^{\prime }\,\tilde{P}%
(x^{\prime }|x)p(y^{\prime }|x^{\prime })\log \frac{P(x^{\prime
}|x)p(y^{\prime }|x^{\prime })}{\gamma ^{1/2}q(y)}  \notag \\
&=&\int dx^{\prime }\,\tilde{P}(x^{\prime }|x)S(x^{\prime })-\int dx^{\prime
}\,\tilde{P}(x^{\prime }|x)\log \frac{P(x^{\prime }|x)}{\gamma ^{1/2}}=%
\mathcal{S}_{\max }~.
\end{eqnarray}%
Therefore any $\tilde{P}(x^{\prime },y^{\prime }|x)$ satisfying the same
constraints as $P(x^{\prime },y^{\prime }|x)$ lowers the relative entropy.

\end{document}